# Scenario Generation of Wind Farm Power for Real-Time System Operation

Trevor Werho, *Member, IEEE*, Junshan Zhang, *Fellow, IEEE* and Vijay Vittal, *Life Fellow, IEEE*, Yonghong Chen, *Senior Member, IEEE*, Anupam Thatte, *Member, IEEE,* Long Zhao

*Abstract*— This work proposes a method of wind farm scenario generation to support real-time optimization tools and presents key findings therein. This work draws upon work from the literature and presents an efficient and scalable method for producing an adequate number of scenarios for a large fleet of wind farms while capturing both spatial and temporal dependencies. The method makes probabilistic forecasts using conditional heteroscedastic regression for each wind farm and time horizon. Past training data is transformed (using the probabilistic forecasting models) into standard normal samples. A Gaussian copula is estimated from the normalized samples and used in real-time to enforce proper spatial and temporal dependencies. The method is evaluated using historical data from MISO and performance within the MISO real-time look-ahead framework is discussed.

*Index Terms*—Copula, MISO, power system, probabilistic forecast, real-time, regression, scenario generation, wind farm.

## I. INTRODUCTION

WIND power has seen major worldwide growth in recent years with more than 60.4 GW installed in 2019 [1].

Wind turbines generate power based on wind speed and do not regulate their output in the same manner as conventional generating units, which leads to increased uncertainty during power system operation [2]. For example, the Midcontinent Independent System Operator (MISO) power system contains an installed wind capacity of over 26 GW and wind forecasting error represents a large source of uncertainty in the system. To help manage uncertainty in the system, MISO has utilized a Look-Ahead commitment (LAC) tool, which informs the commitments of fast-start units during real-time operations. The tool executes every 15 minutes to solve the deterministic security constrained unit commitment problem and considers a horizon of 3-hours into the future [3]. More recently, advanced stochastic optimization methods are being developed to solve the economic dispatch (ED) and unit commitment (UC) problems [4]. However, in order for these advanced methods to be utilized, the future uncertainty must be represented in the form of a scenario-based forecast. This work considers the problem of generating scenarios for a large fleet of wind farms in real-time to support such stochastic optimization methods.

Many scenario generation methods are derived from probabilistic forecasts, which can be produced using a variety of stochastic methods such as regression [5], Markov chains [6], or neural networks (NN) [7]. The challenge is then accounting for the temporal (if more than one time step is considered) and/or spatial (if more than one wind farm is considered) dependence between multiple distributions when sampling. This is often accomplished using Copula theory. From early applications of this concept, quantile regression is used to create probabilistic forecasts of aggregate day-ahead wind power production while temporal dependence is captured by a Gaussian copula [8]. In [9], wind speed data is fit to Weibull distributions; temporal dependence is maintained using an ARMA model while a Gaussian copula is used to model spatial dependence. Other copulas have also been used, such as [10], which produces probabilistic forecasts using a hybrid NN and quantile regression approach but uses a Gumbel copula to better fit the dependence asymmetries in wind power data. Both [11] and [12] use a vine Copula which allow different bivariate copula shapes to be used throughout the spatial/temporal dependence structure. In [13], fewer and wider spaced time steps are considered in day-ahead scenario generation. The probabilistic forecasts are then assumed to be independent allowing scenarios to be sampled by selecting quantiles of interest and not requiring Monte Carlo sampling. Then there are generative adversarial network (GAN) methods that do not rely on any explicit probabilistic forecast, but instead train a NN to produce new trajectories that cannot be distinguished from historic data. GANs have been applied to generate scenarios for day-ahead aggregate [14] and multi-site [15, 16] wind power problems.

This work presents an efficient and scalable method to produce scenarios for a large system of wind farms. This method utilizes probabilistic forecasts using a conditional heteroscedastic model, and spatio-temporal dependencies are captured using a Gaussian copula. The literature on scenario generation is largely focused on the day-ahead time horizon, whereas this work considers the real-time applications. Although there is much overlap between the two problems, there are major differences that arise from the change in time scale. Most notably, real-time applications operate under much tighter time constraints. Consider that the MISO LAC tool executes every 15 minutes, and the UC optimization problem must be solved within 5 minutes. The scenarios that correspond to the present

This work was supported by the United States Department of Energy under the Grant DE-AR0000696 and by the Defense Threat Reduction Agency under the Grant HDTRA1-13-1-0029.

T. Werho, J. Zhang, and V. Vittal are with the School of Electrical, Computer, and Energy Engineering, Arizona State University, Tempe, AZ 85281 USA (e-mails: twerho@asu.edu, junshan.zhang@asu.edu, vijay.vittal@asu.edu)

Y. Chen, A. Thatte, and L. Zhao are with the Midcontinent Independent System Operator, Carmel, IN 46032 USA (emails: ychen@misoenergy.org, athatte@misoenergy.org, lzhao@misoenergy.org)

system conditions need to be generated before the optimization problem can even begin. It can be challenging to produce an adequate number scenarios for a large fleet of wind farms in the available time frame. Execution time is often not discussed in day-ahead studies, but in [9], it is reported that 1,000 scenarios can be generated for a system of 5 wind farms in 3 minutes. Such run times would be of no consequence in a day-ahead study but could pose a problem for real-time. This work presents a framework to generate scenarios that requires minimum online computation, ideal for real-time applications. This work also considers both a spatial and temporal correlation structure within the system scenarios. It is discussed that while other work (using day-ahead data) found that a Gaussian copula was not a sufficient match, it is shown (using real data) that a Gaussian copula is effective at capturing the spatio-temporal dependencies within these short time horizons. Finally, this work finds that using a stochastic model to update an NWP forecast in real-time, considering the most recent farm measurements and past NWP behavior, can result in improved forecast performance.

Section II discusses the details of the proposed scenario generation model. Results from applying the model to the MISO system data are presented in Section III, and conclusions are drawn in section IV.

## II. SCENARIO GENERATION

This section presents the details of the scenario generation method. This work considers the effort of generating scenarios for multiple wind farms and multiple time horizons. Here, wind farms ($w$) will be numbered from $w = 1$ to $w = n_w$ and the specific look-ahead horizon ($\tau$) will be numbered from $\tau = 1$ to $\tau = n_\tau$.

### A. Probabilistic Forecasts using a Conditional Heteroscedastic Model

Probabilistic forecasts are made for each wind farm and each look ahead horizon. With $n_w$ wind farms and $n_\tau$ look-ahead horizons, there are $n_w * n_\tau$ probabilistic forecasting models in this application. Each model focuses on a single look-ahead time for a single wind farm. Probabilistic forecasts are made using a conditional heteroscedastic model. This model makes the assumption that the distribution of forecast errors is constant in shape, but the variance of the distribution is conditional on the independent explanatory variables. The model is

$$y_t = f_1(X_t) + e_t \quad (1)$$

where, $y_t$ is the dependent response variable, $X_t$ is the set of all independent explanatory variables, $f_1(\bullet)$ is a function that maps the explanatory variables to the response variable, and $e_t$ is the error term such that,

$$e_t = u_t h_t \quad (2)$$

Here, $u_t$ is the random error at time $t$, randomly sampled from an unknown distribution, and $h_t$ is a scaling factor equal to or greater than zero. In the heteroscedastic model, $h_t$ is assumed to be dependent on the explanatory variables such that

$$h_t = f_2(X_t) \quad (3)$$

where, $X_t$ is the set of all relevant independent explanatory variables and $f_2(\bullet)$ is a function that maps the explanatory variables to the scaling factor.

The underlying error distribution is unknown but can be learned with sufficient training data. The estimated value of the dependent variable at time $t$, ($\hat{y}_t$) can be written as

$$\hat{y}_t = f_1(X_t) \quad (4)$$

Then, using (1) and (2), we have that

$$y_t = \hat{y}_t + u_t h_t \quad (5)$$

Rearranging terms gives

$$u_t = \frac{y_t - \hat{y}_t}{h_t} \quad (6)$$

Finally,

$$u_t = \frac{y_t - f_1(X_t)}{f_2(X_t)} \quad (7)$$

Therefore, using training data ($y$ and $X$) and the forecast functions ($f_1(\bullet)$ and $f_2(\bullet)$), one sample from the unknown error distributions ($u_t$) is also known. By transforming all the training data, the distribution can be learned empirically. The empirical CDF can be found using

$$\hat{F}_{w,\tau}(z) = Pr(z < u) = \frac{1}{n}\sum_{t=1}^{n} 1_{u_t > z} \quad (8)$$

where, $\hat{F}_{w,\tau}(z)$ is the empirical CDF of the forecasting model for wind farm $w$ and look-ahead horizon $\tau$ and $1_{u_t > z}$ is equal to 1 if the event $u_t > z$ is true [17].

This method is flexible in that there are many ways to define the functions $f_1(\bullet)$ and $f_2(\bullet)$. This is equivalent to a supervised learning input-output mapping problem. This topic has been extensively studied in the literature and include methods of statistical fitting and machine learning architectures [18]. In this work, the functions were implemented using linear regression for its computational efficiency and basis transformations were used to account for non-linearities in the data [19]. Using regression, (1) can be represented as

$$y = \alpha_1 x_1 + \cdots + \alpha_n x_n \quad (9)$$

where $y$ is the response variable, $x_i$ are the independent explanatory variables, and $\alpha_i$ are the regression coefficients. The values of $y$ and $x$ are defined by historical measurements. With the coefficients being the only unknowns, they are estimated using the method of least squares [20]. Equation (3) is slightly different during training. The value of $h_t$ is not directly measured, instead, it is defined as

$$h_t = |y_t - \hat{y}_t| \quad (10)$$

then (3) becomes

$$|y_t - \hat{y}_t| = \beta_1 x_1 + \cdots + \beta_n x_n \quad (11)$$

With only the regression coefficients ($\beta_i$) unknown, they can be estimated using the method of least squares. The scaling factor ($h_t$) is used to describe the size of distributions relative to each other, conditioned on the inputs ($x_i$). Because of this, $h$ is scale invariant. That is to say, if the estimation of $h$ is adjusted by a constant factor of 0.5 then the estimation of the distribution $u$ will be 2 times too wide. However, when the method is used in practice the two terms multiply and the error in estimation is cancelled out.

Regardless of the method used, the features (independent explanatory variables) and the target (dependent response variable) need to be defined. For this application, measurements that were consistently and reliably available for all wind farms were the wind farm power outputs. Also available is the real-time wind forecast used by MISO, which is an NWP forecast.

Weather information, such as temperature, was not available for this study. The features that were found to be most useful were:
- The NWP forecast created at time $t$ and looking ahead to horizon $\tau$ ($F_t^\tau$): The NWP forecast values trailing the target horizon were also useful ($F_t^{\tau-i}$).
- The wind farm power output ($P_t$): useful for all time horizons, however, only the past several measurements contributed to improved forecast performance.
- Past NWP forecasting errors ($E_{t-\tau}^t = P_t - F_{t-\tau}^t$).
- Past NWP forecasting errors from neighboring wind farms were also incorporated and provided a modest performance improvement.

For this problem, the target was chosen to be the future NWP forecasting error or $E_t^\tau = P_{t+\tau} - F_t^\tau$.

*B. Estimating the Gaussian Copula*

One major challenge of generating scenarios is to produce scenarios with realistic behavior. Temporal dependencies between adjacent time steps determine the ramping behavior of the wind farms. Spatial dependencies determine how the generation of multiple wind farms change together. There can even be significant spatio-temporal dependencies if one farm typically lags (or leads) behind another. Also, these dependencies become more complex as more wind farms and time steps are included within the scenario. An effective method for capturing these relationships is with the use of a copula. A copula describes the joint distribution between multiple random variables. It is a powerful tool because it decouples the estimation of marginal distributions and their dependence structure. In [8], a Gaussian copula was used to capture the temporal dependencies in scenarios using day-ahead wind power data. In [9], a Gaussian copula was used to capture the spatial dependencies between 5 wind farms in scenarios using wind speed data. The bivariate Gaussian copula has a very symmetric football-like shape. However, [21] found significant asymmetries within spatial dependencies when using wind power data and recommended the use of a Gumbel copula. Later works [10, 11, 12] have used other copula shapes to better fit these asymmetries. In the problem considered in this work, there are both multiple wind farms and time horizons. However, it was found that a Gaussian copula provides a good estimation of both the temporal and spatial dependencies in this application. Examples of the dependency structure are shown in Section III.

The Gaussian copula can be implemented by constructing a standard Gaussian multivariate random variable [8]. This variable is defined by a simple correlation matrix ($\Sigma_N$), which effectively enforces both the spatial and temporal dependencies during scenario sampling and is easily estimated from the training data. Using (7), the raw measurements ($y$ and $X$) are transformed into random samples from the underlying error distribution ($u_t$). Using these measurements, the CDFs for all $n_w * n_\tau$ probabilistic forecasting models can be estimated using (8). In general, the distributions will not be Gaussian. In order to utilize non-Gaussian distributions in this approach, the random samples from the data need to be transformed using

$$g_t = F_G^{-1}[F_{w,\tau}(u_t)] \qquad (12)$$

where, $F_G^{-1}$ is the inverse CDF function for a standard Gaussian distribution and $g_t$ is the transformed sampled [9]. The transform in (12) effectively adjusts the sample $u_t$ such that it came from a Gaussian distribution. The process can also be reversed by using

$$u_t = F_{w,\tau}^{-1}[F_G(g_t)] \qquad (13)$$

Reference [9] has a detailed depiction of this process. The covariance matrix is then estimated using

$$\Sigma = \frac{1}{n-1}\sum_{t=1}^{n}(X_t - E[X_t])(X_t - E[X_t])^T \qquad (14)$$

Where $\Sigma$ is the covariance matrix, $E[\bullet]$ is the expected value operator, $X_t$ is the vector containing the transformed samples from all models for time $t$ ($X_t = [g_t^{1,1}, g_t^{1,2}, \ldots, g_t^{n_w,n_\tau}]$), and $\bullet^T$ denotes the transpose. The correlation matrix can be obtained by normalizing the covariance matrix as

$$\Sigma_N = \Sigma \oslash (\sigma\sigma^T) \qquad (15)$$

where, $\Sigma_N$ is the correlation matrix, $\oslash$ denotes element-by-element division, and $\sigma$ is a vector of the square roots of the diagonal elements of $\Sigma$ [8].

*C. Model Training and Scenario Sampling*

This section presents the procedure to train models and sample scenarios for the system and is discussed step by step. The procedure is labeled as parts that can be done offline, and parts that must be done online.

1. (Offline): Wind farm data is collected, and a probabilistic forecasting model is created for each wind farm and time horizon. The features and target for each model are chosen and $f_1(\bullet)$ and $f_2(\bullet)$ are fit using the training data.

2. (Offline): Using (7), the wind farm measurements are transformed into random samples from the model error distribution ($u_t$).

3. (Offline): The empirical CDF for each model is estimated using (8).

4. (Offline): The transformed samples ($u_t$) are converted to Gaussian ($g_t$) using (12).

5. (Offline): The covariance matrix ($\Sigma$) is estimated using (14) and then the correlation matrix ($\Sigma_N$) is estimated using (15). A standard Gaussian multivariate random variable is defined using the correlation matrix ($\Sigma_N$) and a vector of means ($\mu$) containing all zeros.

6. (Offline): The standard Gaussian multivariate random variable is sampled in a Monte Carlo fashion. A single sample from this variable is a vector that contains $n_w * n_\tau$ values (one for every wind farm and time step within the problem). These standardized samples ($\hat{g}_t$) represent the underlying noise within the scenarios with respect to the probabilistic model. Notice that these samples are independent of the present system condition. This allows the Gaussian variable to be sampled offline. Sufficient samples should be generated to match the desired number of scenarios.

7. (Offline): The standard samples ($\hat{g}_t$) (from step 6) are converted from Gaussian distributions to the distributions trained by the probabilistic forecasting models ($\hat{u}_t$) using (13).

8. (Online): Read in the most recent measurements and estimate the point forecast ($\hat{y}_t$) and the scaling factor ($\hat{h}_t$) for every probabilistic model using (4) and (3) respectively. These values depend on the most recent measurements from the wind farm, and as such, must be computed online.

9. (Online): The final sampled scenarios are constructed by adjusting the standard transformed samples ($\hat{u}_t$) to account for the current system conditions. This is accomplished by a simple scaling and shifting using

$$\tilde{y}_t = \hat{y}_t + \hat{u}_t \hat{h}_t \tag{16}$$

where $\hat{y}_t$ and $\hat{h}_t$ are determined by the forecasting model and $\tilde{y}_t$ is the final sampled value.

The proposed approach is very efficient and can support real-time applications for very large systems. Notice that only 2 of the 9 steps require online computations. The problem considered in this work has 152 wind farms and 36 look-ahead time horizons. Even so, the proposed approach can generate 1000 scenarios for this system in approximately 10 seconds. The algorithm was implemented in python 3.7 as a single thread program and was run on a 2.2 GHz Intel® Xeon® E5-2699 v4 CPU. This is in contrast with [9] which reported a required 3 minutes to generate 1000 scenarios for only 5 sites and 24 time horizons. The efficiency of the method is two-fold, using the conditional heteroscedastic model, the CDF functions for each model can be constructed offline (step 3). This is different than other methods such as quantile regression, where every quantile of the CDF is dependent on the most recent farm measurements. This requires the CDF to be constructed online, increasing the algorithm complexity. Further, the estimation of the point forecast and scaling factor for each model (step 8) applies to *all* scenarios. Hence, the algorithm scales very well with respect to the number of scenarios desired. For example, increasing the number of generated scenarios by a factor of 10 only corresponds to about 30% additional run time.

## III. PERFORMANCE EVALUATION

The data used in this work comes from the MISO power system. It included 152 wind farms with a maximum output of 15.5 GW. The forecast horizon aligned with the MISO LAC tool which looks ahead 3 hours. The MISO data is at a 5-minute resolution, therefore, the scenarios will contain 36 time steps to span the 3 hour horizon. Wind farms ($w$) will be numbered from $w = 1$ to $w = 152$ and the specific look-ahead horizon ($\tau$) will be numbered from $\tau = 1$ (5 minutes ahead) to $\tau = 36$ (3 hours ahead). The method, presented in Section II, is used to generate scenarios for the MISO system in this section, and its performance is evaluated. Scenarios were generated by following the procedure presented in Section II, C. The probabilistic forecasting functions (step 1) were trained with the most recent 4 weeks of measurements, to account for seasonal changes in wind farm behavior and MISO forecasting procedure. Approximately 3 months of recent data was used to estimate both the empirical CDFs (step 3) and the correlation matrix (steps 4 and 5).

Fig. 1 and 2 show example scenarios created for a single time period. Both figures show 15 equally weighted scenarios depicting the possible future trajectories of wind over the coming 3 hours. Also shown is the future realization and the point forecast made by the model. Fig. 1 shows the aggregation of all individual farm scenarios (and actuals), while Fig. 2 shows only the scenarios generated for a single wind farm. In Fig. 2, by comparing the scenarios to the actual measurements, it is apparent that the model accurately captures the temporal dependencies for this farm. Further, in Fig. 1, the aggregation of all the individual farm scenarios mimics the behavior of the aggregate wind farm power, indicating that the spatio-temporal dependence structure is correctly replicated by the model.

In [8], probabilistic forecasts are also used in conjunction with a Gaussian copula, but applied to day-ahead aggregate system data. In [13], the authors commented that the scenarios generated in [8] contained unrealistic ramping behavior that was visually inconsistent with the future realization. This was likely due to the availability of data to estimate the correlation matrix. With hourly day-ahead data, one year of data only contains 365 days. However, in this real-time application, one year contains over 100,000 samples. The correlation matrix used in this study was estimated with approximately 4000 samples from the prior 3 months before operation. With the increased availability of training data, the issues seen in [8] were avoided.

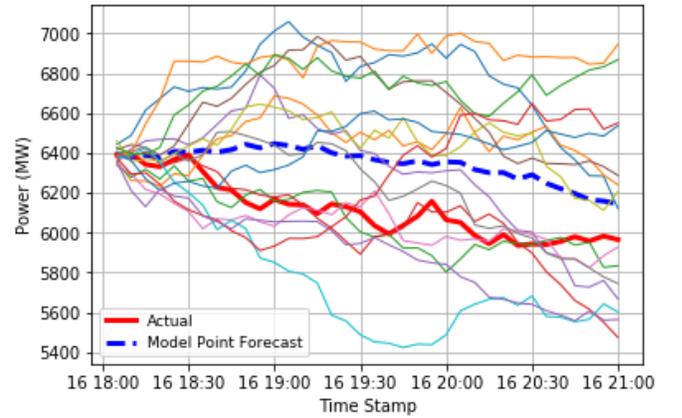

Fig. 1. Example scenarios for the aggregate wind generation. The figure includes the future actual realization (bold red line), the point forecast made by the model (dashed blue line), along with 15 equally weighted scenarios.

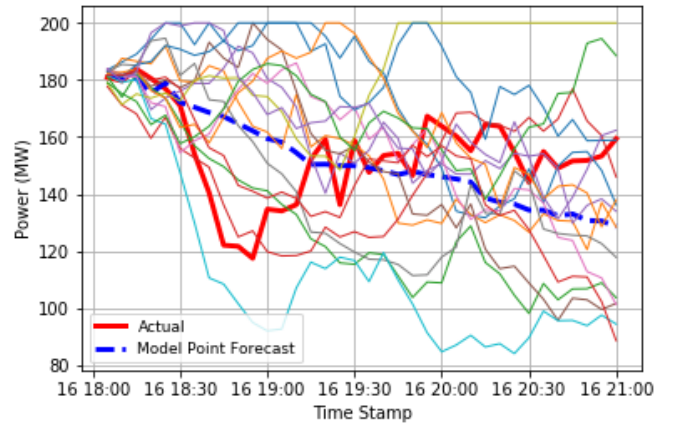

Fig. 2. Example scenarios for a single wind farm. The figure includes the future actual realization (bold red line), the point forecast made by the model (dashed blue line), along with 15 equally weighted scenarios.

### A. Point Forecast Performance

Although a point forecast is not explicitly considered in stochastic optimization problems, the point forecast is an important aspect of the model. The point forecast made by the stochastic model, described in Section II, was compared to the existing MISO NWP forecast to understand if it can provide any benefit to the existing MISO real-time forecasting procedure.

Forecasting performance was evaluated using the root-mean-squared-error (RMSE) metric, which penalizes large errors much more than small errors. The RMSE is defined as

$$RMSE = \sqrt{\frac{1}{N_t}\sum_t (P_t - \hat{P}_t)^2} \quad (17)$$

where, $P_t$ is the measured power output, $\hat{P}_t$ is the forecasted power output, and $N_t$ is the number forecasts considered. Forecasting was conducted over several periods throughout the year, ranging from single days to a week in length, executing every 15 minutes.

The point forecast made by the model was able to consistently provide lower forecasting errors than the existing MISO forecast for all time horizons, with the largest improvements seen in the early horizons. This observation was consistent with both the aggregation of all wind farms and when focusing on the performance of only a single farm. Described in Section II, the proposed forecasting model effectively adjusts the given MISO forecast based on recent past measurements and past MISO forecasting performance. This indicates that the MISO forecasting procedure could benefit from using such a strategy. However, it is not yet clear if the magnitude of improvement would result in a significant benefit (cost savings or improved reliability) if used in real-time system operation. Unfortunately, the NWP forecast used by MISO is sensitive data and quantitative results cannot be made available.

### B. Probabilistic Forecast Reliability

Another important aspect of the model is the probabilistic forecast. Recall from Section II, probabilistic forecasts are made using a conditional heteroscedastic model. Fig. 3 shows an example CDF that was estimated from the data for a single wind farm, considering forecasts 60 minutes into the future. Also shown is a standard Gaussian distribution for comparison. It is clear that the distribution of errors estimated by the model is non-normal, with larger probabilities in the tails of the distribution.

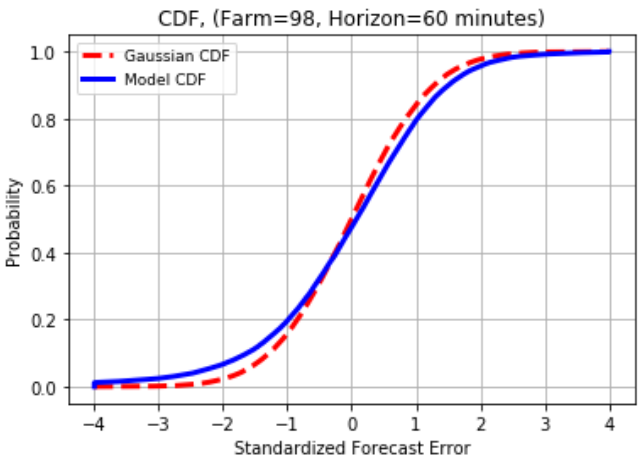

Fig. 3. Example CDFs for a wind farm looking 60 minutes ahead. The figure shows the CDF estimated by the model (solid blue line) and the CDF for a standard Gaussian distribution (dotted red line).

It is important to know if the CDF estimated by the model, using past data, is valid when applied to future data. This can be accomplished by creating a reliability diagram. A reliability diagram shows how consistent the prediction intervals of a forecast distribution are when compared to the actual observations. For example, when considering an event that is forecasted to have a 10% probability of occurring, is should be observed 10% of the time in the future, to be a reliable forecast. The probabilistic forecasting model was trained using approximately 3 months of data (June, July, August) and was tested using the following 3 months (September, October, November). Fig. 4 shows the reliability diagram for the model $w = 98$ and $\tau = 12$. The figure shows the forecast reliability for the forecasting model (solid blue) and the reliability if the forecast is assumed to be Gaussian (dotted red). From Fig. 4, the model clearly makes a reliable forecast for this wind farm and horizon. Similar observations were made using the other combinations. Also shown in Fig. 4 is the reliability for a normal distribution, which does not quite match the actual measurements. That being said, assuming a Gaussian distribution is not a terrible assumption in this application and could be a reasonable approximation if the algorithm implementation effort is a major hurdle. This is contrary to day-ahead forecasting which can have non-normal distributions of error with much higher degree, as discussed in [13].

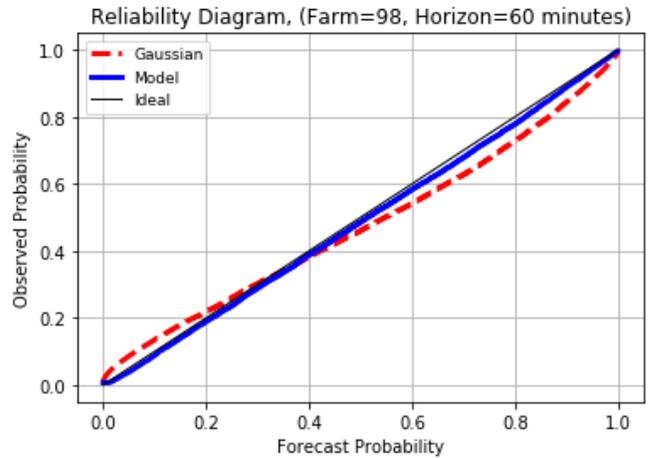

Fig. 4. The reliability diagram for the forecasting model (farm 98, looking 60 minutes ahead). The reliability of the model distribution is shown in solid blue while the reliability of a Gaussian distribution is shown as a dotted red line.

### C. Copula Evaluation

Just as the marginal distributions estimated by the model can be compared to the real data, the joint distributions estimated by the copula are evaluated here in a similar manner. The copula structure can be visualized by representing the data in the rank domain. The data is first transformed into forecasting errors ($u_t$) using (7). The data is converted to the rank ($r_t$) using

$$r_t = F_{w,\tau}(u_t) \quad (18)$$

where, $r_t$ is a uniform random variable over the interval [0,1]. This transformation decouples the dependency structure from the changing system conditions [21].

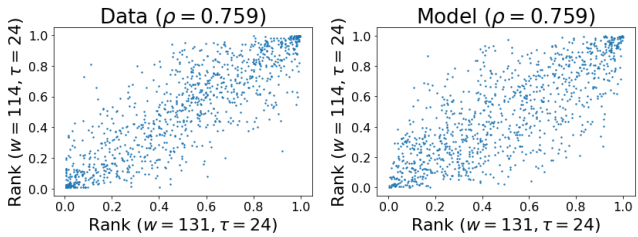

Fig. 5. The spatial correlation between two wind farms, using real data (left) and sampled from the model (right).

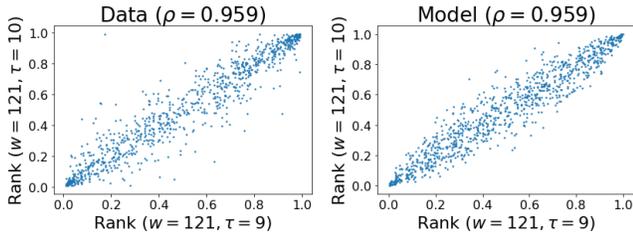

Fig. 6. The temporal correlation between adjacent time steps for a single wind farm, using real data (left) and sampled from the model (right).

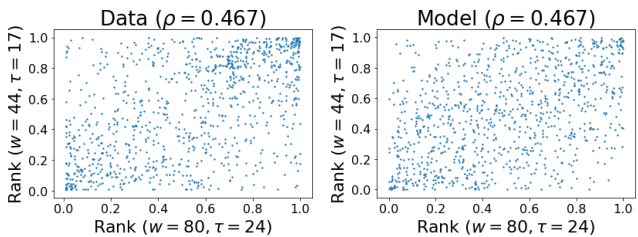

Fig. 7. The spatial-temporal correlation between two wind farms and horizons 35 minutes apart, using real data (left) and sampled from the model (right).

Figs. 5, 6, and 7 show the dependency structure between a few combinations of wind farms and time horizons. On the left of each figure is a scatter plot of real data in the rank domain. The Pearson correlation coefficient ($\rho$) that was estimated from this data is shown. On the right of each figure is a scatter plot of data sampled from a Gaussian copula using the same value of $\rho$. Fig. 5 shows the spatial correlation between two wind farms, both using the 2 hour ahead time horizon. Fig. 6 shows an example of temporal correlation between two adjacent time steps for the same wind farm. Finally, Fig. 7 shows the spatio-temporal correlation between two wind farms 35 minutes apart. It can be seen from the figures that the Gaussian copula is a good approximation to the dependence structure in this application. The structure using the real data has a very symmetric shape along both diagonals, even for spatial dependencies. This is different than what was concluded in [21] using day-ahead wind power data. This is likely a result of the shorter timescale and that NWP forecasts are much more accurate over these shorter time horizons.

### D. Aggregate Scenario Performance

The approach presented here produces scenarios while representing each individual wind farm in the system. If the spatio-temporal dependencies are captured well, then the aggregation of the scenarios for individual farms should return viable scenarios for the aggregate wind power. An example of this can be seen in Fig. 1, however, this is further explored in this section. An alternative way to produce scenarios for the aggregate wind power is to first aggregate all the wind data and create a model that only considers the aggregate. A comparison of these two strategies is shown here. The first strategy is the proposed method, representing each individual wind farm, as discussed in Section II. In the second, all farm data is aggregated, and the proposed model is applied to the aggregate (1 site, 36 time horizons). The two approaches were evaluated over a 6 day period in the fall. Every 15 minutes, 200 equally weighted scenarios were sampled. The quality of the aggregate scenarios was evaluated using a variety of performance metrics.

The Energy score is a proper scoring rule where a perfect forecast results in the best score. This metric quantifies both the accuracy and reliability of the scenarios. The Integrated Distance metric is a simple metric that evaluates the sum of the absolute distance between the realization and all scenarios. Finally, the Variogram score is a proper scoring rule that considers pairwise differences between components to evaluate the multivariate correlations of the data. All scores are negatively oriented, where lower scores indicate better performance. Additional details of these metrics are discussed in [22] and [13].

Table I shows the performance metrics for both approaches over the test period. Over this interval, the two methods are similar, but the method considering every individual wind farm generated consistently better aggregate scenarios. This further illustrates that the Gaussian copula can effectively capture the spatio-temporal dependencies found in this application.

Fig. 8 and 9 show examples of scenarios generated using the two different levels of representation for a single time period. Both the future realization and the model point forecast are shown, along with 15 equally weighted scenarios. The similarities between the two are apparent.

TABLE I
PERFORMANCE METRICS EVALUATING THE QUALITY OF AGGREGATE SCENARIOS

|  | Energy | Integrated Distance | Variogram |
|---|---|---|---|
| All Farms Represented | 2058 | 17860 | 149548245 |
| Aggregate Data Only | 2154 | 18142 | 156916887 |

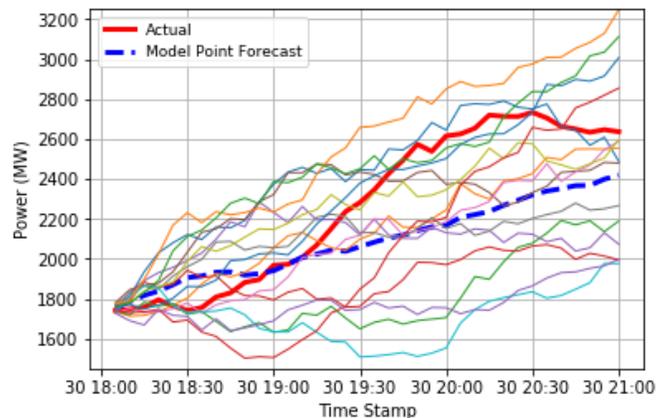

Fig. 8. Example scenarios for the aggregate wind generation, using the proposed model that represents every individual wind farm.

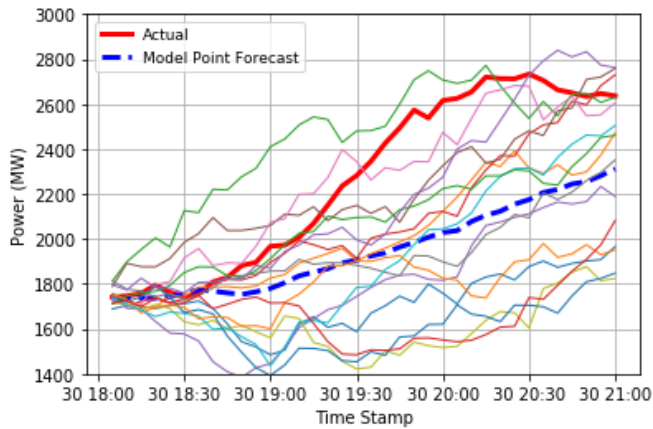

Fig. 9. Example scenarios for the aggregate wind generation using the aggregate model.

## IV. Conclusions

Real-time power system operation is becoming increasingly challenging with the increased presence of uncertainties within the system. Stochastic optimization planning tools are one possible solution to this challenge. However, using these advanced tools requires detailed scenarios that represent possible future system conditions. The work presented here offers an effective method of producing high-quality scenarios for this endeavor. While most work in this area focuses on day-ahead time scales, the method discussed here is well suited for real-time applications. The method is extremely efficient and scales well with large problems. The proposed procedure allows for minimal online computations. This allows the algorithm to execute in a timeframe that does not put pressure the optimization solver and allowing additional scenarios to be considered in the problem.

This study had the opportunity to study actual measurements from the MISO power system. It was found that a Gaussian copula is capable of capturing the spatio-temporal dependencies within this data, and the scenarios generated by the proposed method match the behavior of the actual measurements. Also available in this study was the MISO real-time forecast. It was discovered that using the proposed stochastic forecasting method, to adjust the MISO forecast by considering available measurements, was able to provide less forecasting error than the existing forecasting procedure. However, whether this would translate into tangible benefits is unknown.

The proposed method is flexible and is not only applicable to wind farm data. The same approach has been used to generate scenarios for area load and net imports, also using the MISO system data. Similar results were obtained using these other datasets.

## Acknowledgements

The authors would like to thank MISO for providing the data used in this study. The authors also thank Jessica Harrison, John Harmon, Congcong Wang, Steve Rose, and Jason Howard from MISO for their support and valuable discussions.

**Trevor Werho** (S'14-M'15) received the B.S.E., M.S., and Ph.D. degrees in electrical engineering from Arizona State University, Tempe, AZ, USA, in 2011, 2013, and 2015, respectively. He completed working as a post-doctoral scholar at Arizona State University researching wind and solar forecasting in 2021. His research interests include power system analysis and the integration of renewables into power systems.

**Junshan Zhang** (F'12) received the Ph.D. degree from the School of Electrical and Computer Engineering at Purdue University, West Lafayette, IN, USA, in 2000.
    He joined the Electrical Engineering Department at Arizona State University, Tempe, AZ, USA, in August 2000, where he has been a Professor since 2010. His research interests include communications networks, cyber-physical systems with applications to smart grids, stochastic modeling and analysis, and wireless communications.

Dr. Zhang is a recipient of the ONR Young Investigator Award in 2005 and the NSF CAREER award in 2003. He received the Outstanding Research Award from the IEEE Phoenix Section in 2003.

**Vijay Vittal** (S'78-F'97-LF'21) received the B.E. degree in electrical engineering from the B.M.S. College of Engineering, Bangalore, India, in 1977, the M. Tech. degree from the India Institute of Technology, Kanpur, India, in 1979, and the Ph.D. degree from Iowa State University, Ames, IA, USA, in 1982.

Currently, he is the Ira A. Fulton Chair Professor in the Electrical Engineering Department, Arizona State University, Tempe, AZ, USA.

Dr. Vittal received the 1985 Presidential Young Investigator Award and the 2000 IEEE Power Engineering Society Outstanding Power Engineering Educator Award. He is a member of the National Academy of Engineering.

**Yonghong Chen** (SM'12) received the B.S. degree from Southeast University, Nanjing, China, the M.S. degree from Nanjing Automation Research Institute, China, and the Ph.D. degree from Washington State University, Pullman, WA, USA, all in electrical engineering. She also received the M.B.A. degree from Indiana University, Kelly School of Business, Indianapolis, IN, USA.

She is currently a Consulting Advisor at MISO and leads research and development to address challenges on market design and market clearing system.

She was in the team won 2011 INFORMS Franz Edelman Award for achievement in operations research and management science.

**Anupam A. Thatte** (S'12-M'15) received the B.E. degree in electrical engineering from Pune University, Pune, India, the M.S. degree in electrical and computer engineering from Carnegie Mellon University, Pittsburgh, PA, USA, and the Ph.D. degree in electrical and computer engineering from Texas A&M University, College Station, TX, USA.

He is currently working for the Midcontinent Independent System Operator, Carmel, IN, USA. His research interests include modeling and control of power systems, grid integration of renewable energy, and electric markets.

**Long Zhao** received the B.S. degree in automation engineering in 2006 and the M.S. degree in computer science in 2009 both from Xian Jiaotong University, China, and the Ph.D. degree in industrial engineering from the University of South Florida, Tampa, FL USA, in 2013. He is currently a research and development advisor at the Midcontinent Independent System Operator, Inc. His research interests include mixed-integer programming, robust optimization, stochastic programming and their applications in power systems.